\documentclass [12 point,onecolumn,a4wide]{article}
\usepackage {vpage}
\usepackage {vmargin}
\usepackage {graphicx}
\setpapersize {Afour}
\setmarginsrb {1in}{1in}{1in}{1in}{0in}{.3in}{0in}{.3in}

\begin{document}
\centerline {\Large Logarithmic Singularities of Specific Heat and Related }

\smallskip
\centerline {\Large Properties of Liquid $^4He$ Near $\lambda$-Point}

\vspace{1cm}

\centerline {\bf Simanta C.$ ^{1,2} \quad$ {\bf and Yatendra S. Jain$^1$}}

\smallskip
\centerline{$^1$Department of Physics, North-Eastern Hill University, 
Shillong-793 022, India}

\centerline{$^2$Department of Physics, St. Anthony's College, 
Shillong -793 001, India} 

\begin{abstract}
The singularity of specific heat ($C_p$) and related properties 
({\it viz.} thermal expansion coefficient, $\alpha_p$, compressibility, 
$\kappa_T$, and pressure coefficient, $\beta$) of liquid $^4He$ at 
$\lambda-$point is studied and the accuracy of its logarithmic nature as 
concluded for the first time from a microscopic theory of a system of 
interacting bosons is examined.  A very good agreement between the results 
of this theory and experiments concludes that the singularity 
is intrinsically logarithmic.  However, as shown by other studies 
reporting carefully measured $C_p$ of liquid $^4He$ around $T_{\lambda}$ 
for $|t| = |(T_{\lambda}- T)/T_{\lambda}|$ ranging between $10^{-1}$ to 
$10^{-9}$, in and out of earth's gravitational field and in finite size 
samples, weak effects arising from earth's gravity and small sample size 
round it off and $C_p$ assumes asymptotic nature near $t \approx 0$.  

\end{abstract}

\section{Introduction:}
The $\lambda$-transition of liquid $^4He$ is characterized by logarithmic 
singularity of: (i) specific heat at constant pressure ($C_p$), (ii) expansion 
coefficient ($\alpha_p$), (iii) isothermal compressibility ($k_T$) and (iv) 
pressure coefficient ($\beta$) [1,2] with varying strength.  An 
excessively high $C_p$ at $\lambda$-point indicating its singular behavior 
was first observed in 1932 by Keesom and Clusius [3], while a series of more 
accurate measurements [4-9] later confirmed its logarithmic nature.  
Recently, Lipa and coworkers [10,11] have reported very carefully measured 
$C_p$, {\it in and out of earth's gravitational field}.  To isolate the 
effect of gravity they made their measurements on STS-52 space craft, 
space shuttle Columbia.  While the 
net dependence of their $C_p$ on $t = (T_{\lambda}-T)/T_{\lambda}$ is 
certainly asymptotic but their analysis indicates that once the observed 
$C_p$ is corrected for the factors which round off its values near 
$T_{\lambda}$, the corrected $C_p$ fits closely logarithmic variation. 
Similarly, careful investigation of the effects of finite size of the 
sample performed by Lipa {\it et al} [12-14] and Gasparini {\it et al} 
[15, 16] also reveals that $C_p$ 
variations are intrinsically logarithmic; they assume asymptotic 
dependence on $t$ when the sample size becomes smaller than coherence 
length.

\bigskip
One also finds that early measurements of density [17,18] showing peak at 
$\lambda$-point indicative of a divergence of $\alpha_p$ were repeated with 
improved accuracy by Atkins and Edwards [19], Kerr [20], Edwards [21], and 
Chase {\it et. al.} [22] for detailed analysis.  These results are reviewed 
elegantly by Kerr and Taylor [23].  It is evident that $\alpha_p$ for its 
linear relation with $C_p$ should have logarithmic divergence at 
$\lambda-$point.  Similarly, the divergence of $\beta$ investigated by 
Ahlers [9], Lounasmaa and Kaunisto [24], Lounasmaa [25], and Kiersted 
[26, 27] is ought to be logarithmic.  We note that thermodynamic relations 
between various response functions, as discussed by Rice [28], 
Pippard[29, 30], and Buckingham and Fairbank [6], indicate that $K_T$ is an 
asymptotically linear function of $C_p$ and hence of $\alpha_p$.  
Evidently, $k_T$ is naturally expected to show logarithmic singularity 
at $\lambda-$point.  However, the identification of this behavior 
from experimental observation is obscured by large regular contribution.  

\bigskip
Despite a long and rich history of experimental work and untiring efforts 
of nearly seven decades for developing a viable microscopic theory of liquid 
$^4He$, the exact nature and origin of the above stated singularities has 
been unknown.  The importance of this is reflected by a comment from Feynman 
in his book [31].  Recently, Jain [32] used unconventional approach to 
develop long awaited microscopic theory of a system of interacting bosons 
such as liquid $^4He$ based on macro-orbital representation of a particle 
in a many body system and obtained a relation for the logarithmic singularity 
of $C_p$.  This paper examines his relation for its agreement with 
experiments and uses it to find similar relations for $\alpha_p$, $K_T$ 
and $\beta$.

\bigskip
\begin{figure}
\begin{center}
\includegraphics [angle = -90, width = 0.7\textwidth]{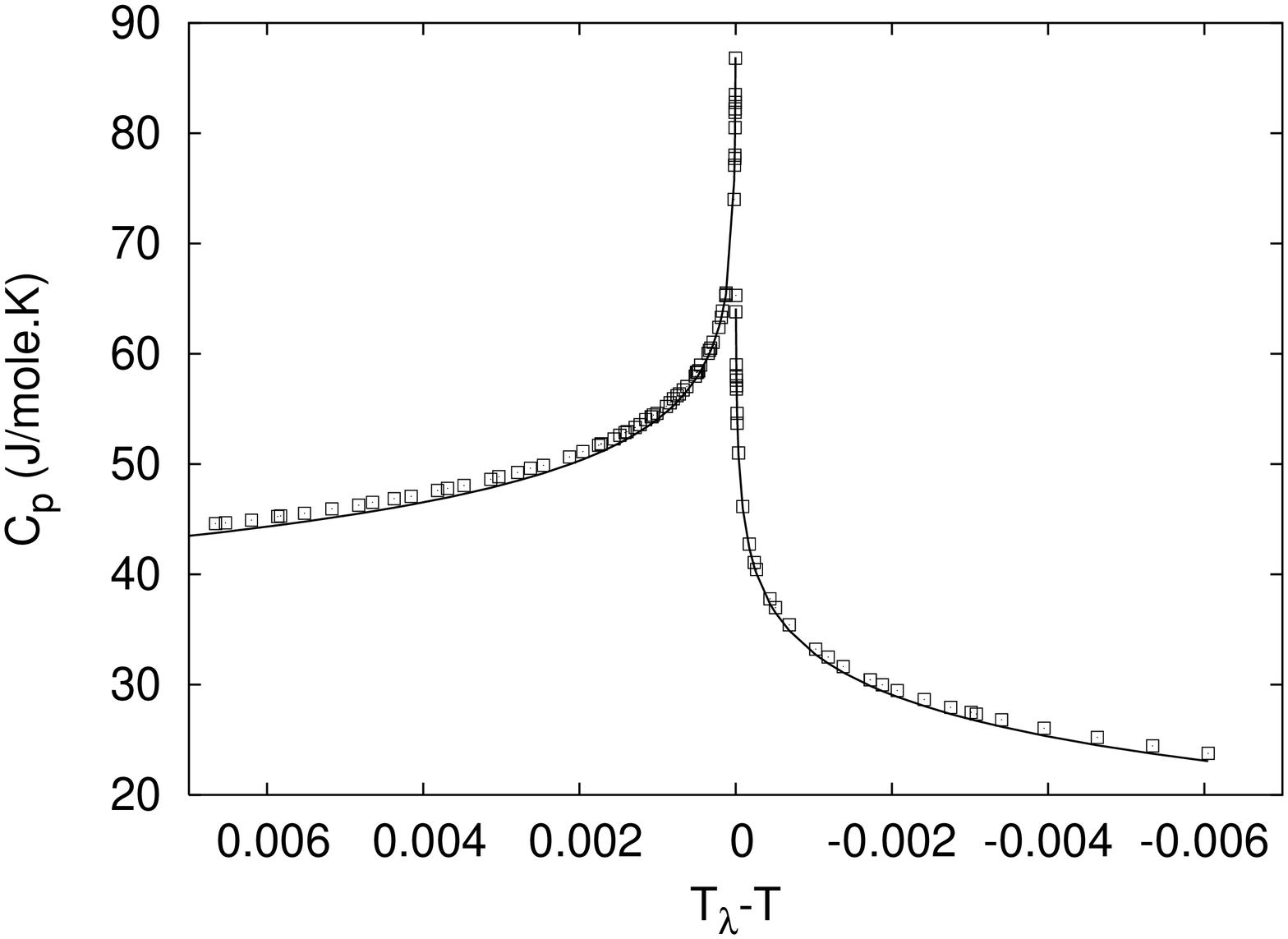}
\end{center}
\bigskip
\noindent
Fig. 1 : $C_p$ of liquid $^4He$ around $\lambda-$point with points indicating 
experimental values from [8] and line representing our theoretical results. 
\end{figure}
\section{Results}
\subsection{Specific heat $C_p$}
\bigskip
For a small range of $|t| <0.1 ^oK$, Jain's theory concludes  

\begin{equation} C_p = -A\ln|t| + B
\end{equation} with

\begin{equation}
A = \frac{N}{T_\lambda}k_BT_o2\nu
\end{equation} and 
\begin{equation}
B = \frac{N}{T_\lambda}k_BT_o[2\ln(\delta\phi_\lambda(0))-\ln4 + 2\nu]
\end{equation}

\bigskip
\begin{figure}
\begin{center}
\includegraphics [angle = -90, width = 0.7\textwidth]{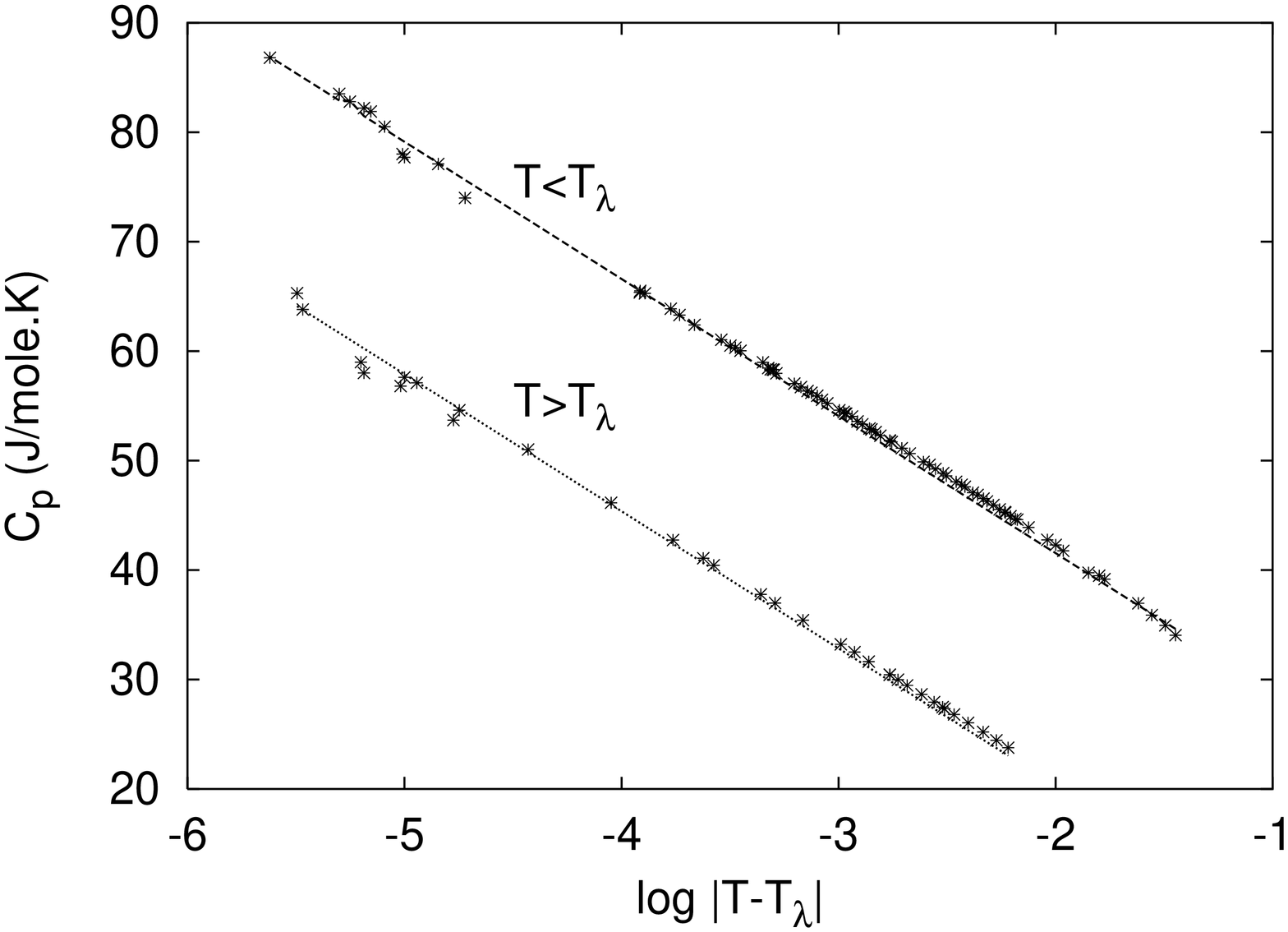}
\end{center}
\bigskip

\centerline {Fig. 2 : $C_p$ $vs.$ $\log|T_{\lambda}-T|$ curve for 
liquid $^4He$ around $\lambda$-point on log scale.}   
\end{figure}  

In order to demonstrate the fact that Eqn. 1 obtained from his 
theory can correctly account for the logarithmic divergence of $C_p$, 
Jain [32] used $T_o=1.49 K$, $\nu = 0.55$ and $\delta\phi_{\lambda}(0) = \pi$ 
to conclude $A = 5.71$ (both for $T < T_\lambda$ and $T > T_\lambda$) 
matching closely with: (i) similar estimates based on Widom-Kadanoff 
scaling laws [30, 33-35] and (ii) $A= 5.1$ for $T < T_\lambda$ and 5.355 
for $T > T_\lambda$ obtained from experimental $C_p$.  While the close 
agreement between theoretical $A$ and experimental $A$ showed the 
accuracy of Eqn. 1, but the choice of parameters rendered $B = - 10.35$ 
which, however, differs significantly from B = -7.77 (for $T>T_\lambda$) 
and 15.52 (for $T<T_\lambda$) estimated from experimental 
$C_p$ [2].  In this paper we examine this aspect more deeply to find 
that: (i) $\delta\phi_{\lambda}(0)$ should, in principle, be lower than 
$\pi$ because the shift in $\phi-$positions of all particles in the 
process of their order-disorder in $\phi$ space need not be equal to 
$\pi$ and (ii) $\delta\phi_{\lambda}(0)$ should not, necessarily, be equal 
for both sides of $T_{\lambda}$ because particles on $T_{\lambda}^+$ side 
nearly have random positions in $\phi-$space while the same on $T_{\lambda}^-$ 
side are largely locked with $\Delta\phi = 2n\pi$ which indicates that 
$\delta\phi_{\lambda}(0)$ on $+ve$ side should be slightly lower than 
$\pi$, while that on $-ve$ side should be fairly small.  Guided by these 
points, we find that $\delta\phi_{\lambda}(0) = 0.75\pi$ for $T>T_\lambda$ 
and $\delta\phi = 0.084\pi$ for $T<T_\lambda$ with $\nu= 0.47$ render 
$A = 5.1148$ and $B = 16.9319$ for $T<T_\lambda$ and $A = 5.1148$ and 
$B = -6.8930$ for $T>T_\lambda$ which agree closely with their 
experimental values.  This agreement can be better perceived from 
Figs. 1-3 where we plot our calculated $C_p$ $vs.$ $T_{\lambda}-T$ along 
with the experimental values [8] for comparison.  Further since $A$ depends 
on $\nu$ and our choice of its equal value for both sides of $T_{\lambda}$ 
renders $A^+ = A^-$ whose experimental values, however, have small 
difference [10, 36, 37].  Evidently, if $\nu$ is presumed to have slightly 
different values on two sides of $T_{\lambda}$ for the same reason, the 
difference in $A^+$ and $A^-$ is easily explained.  Evidently, Jain's theory 
accounts for the observed logarithmic singularity in its all details.  

\bigskip
\begin{figure}
\begin{center}
\includegraphics [angle = -90, width = 0.7\textwidth]{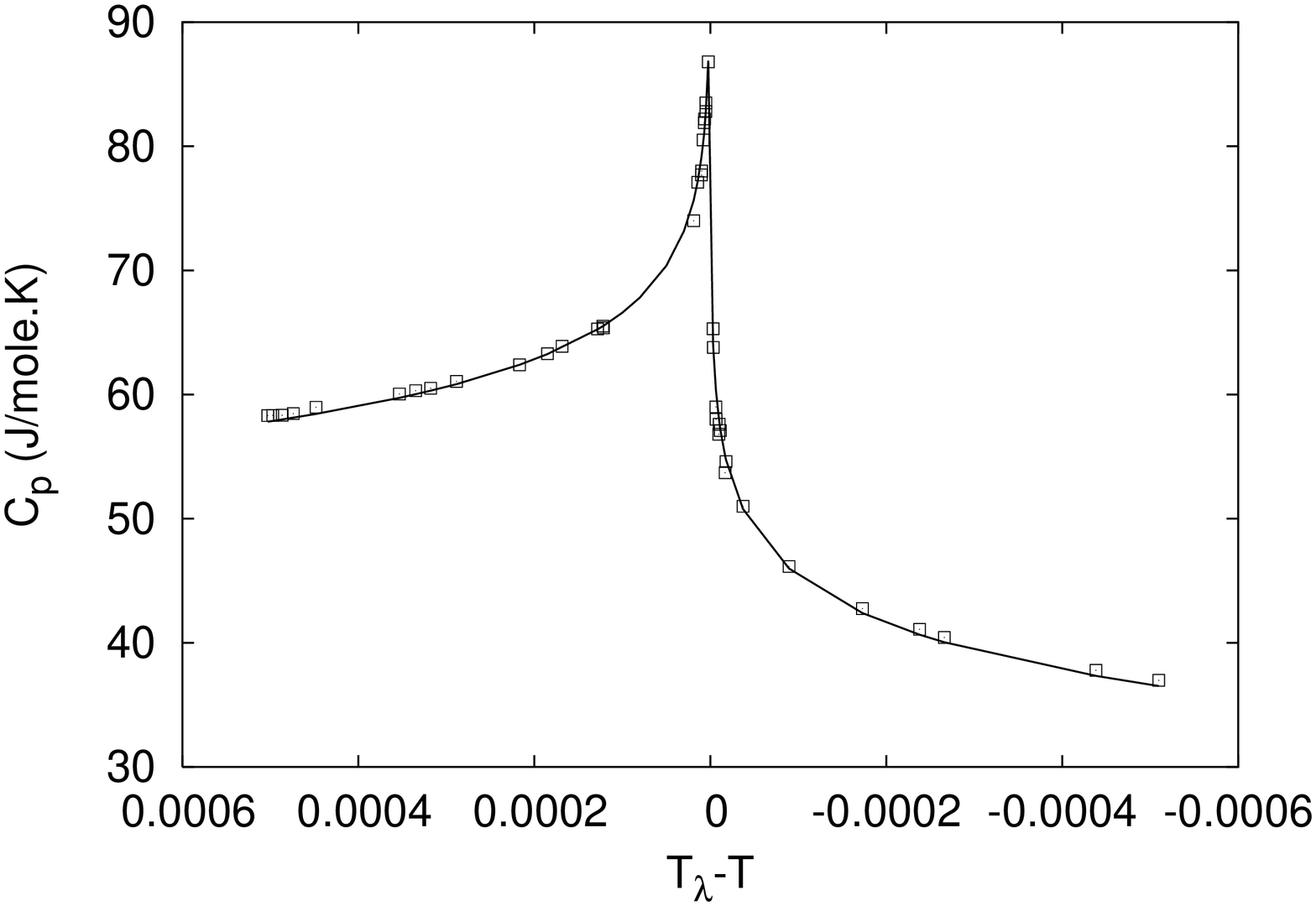}
\end{center}
\bigskip
\noindent
Fig. 3 : $C_p$ $vs.$ $T_{\lambda}-T$ curve for liquid $^4He$ around 
$\lambda$-point (the points in this figure are closer to $\lambda-$point 
in comparison to Fig. 1). Curves on two sides of $\lambda-$point 
join each other ({\it as shown here}) when weak perturbation on 
logarithmic singularity round it off.   
\end{figure} 

\subsection{Expansion Coefficient}
PBF relation between $C_p$ and $\alpha_p$ [2] along an arbitrary 
thermodynamic path renders:
\begin{equation}
C_p = T (\frac{\partial S}{\partial T})_t + VT(\frac{\partial P}{\partial T})_t \alpha_P
\end{equation}
Close to the $\lambda$ line, this relation yields
\begin{equation}
\alpha_P = A_\alpha \log{\left| T-T_\lambda\right|}+B_\lambda
\end{equation}
with
\begin{equation}
A_\alpha = -2.3025 \frac{A}{T_\lambda V_\lambda\left( \partial P/\partial T\right)_\lambda}
\end{equation}
and 
\begin{equation}
B_\alpha = \frac{B + A\ln{T_\lambda}-T_\lambda\left(\partial S/\partial T\right)_\lambda}{V_\lambda T_\lambda (\partial P/\partial T)_\lambda}
\end{equation}
\smallskip
where $A$ and $B$ are the same coefficients of Eqn.1 for $C_p$. 
To estimate the coefficient $A_\alpha$ and $B_\alpha$ we use 
$\left(\partial P/\partial T\right)_\lambda = -112.5$ bar/$^oK$, 
$\left(\partial S/\partial T \right)_\lambda = 102$ cm$^3$ bar/mole/$^oK^2$ 
and $V_\lambda = 27.38$ cm$^3$/mole [2]. 
These parameters give $A_\alpha = 0.0166$ and $B_\alpha = 0.0032$ for 
$T<T_\lambda$ and $A_\alpha = 0.0166 $ and $B_\alpha = 0.0367$ for 
$T>T_\lambda$, while experiments reveal $A_\alpha = 0.01680$ and 
$B_\alpha = 0.0025$ for $T < T_\lambda$ and $A_\alpha = 0.0169$ and 
$B_\alpha = 0.0379$ for $T > T_\lambda$ [1].  We depict our results 
in Fig. 4 along with the experimental results of [19] and [9], 
respectively, marked as Exp.(1) and Exp(2).  

\bigskip
\begin{figure}
\includegraphics [angle = -90, width = 0.9\textwidth]{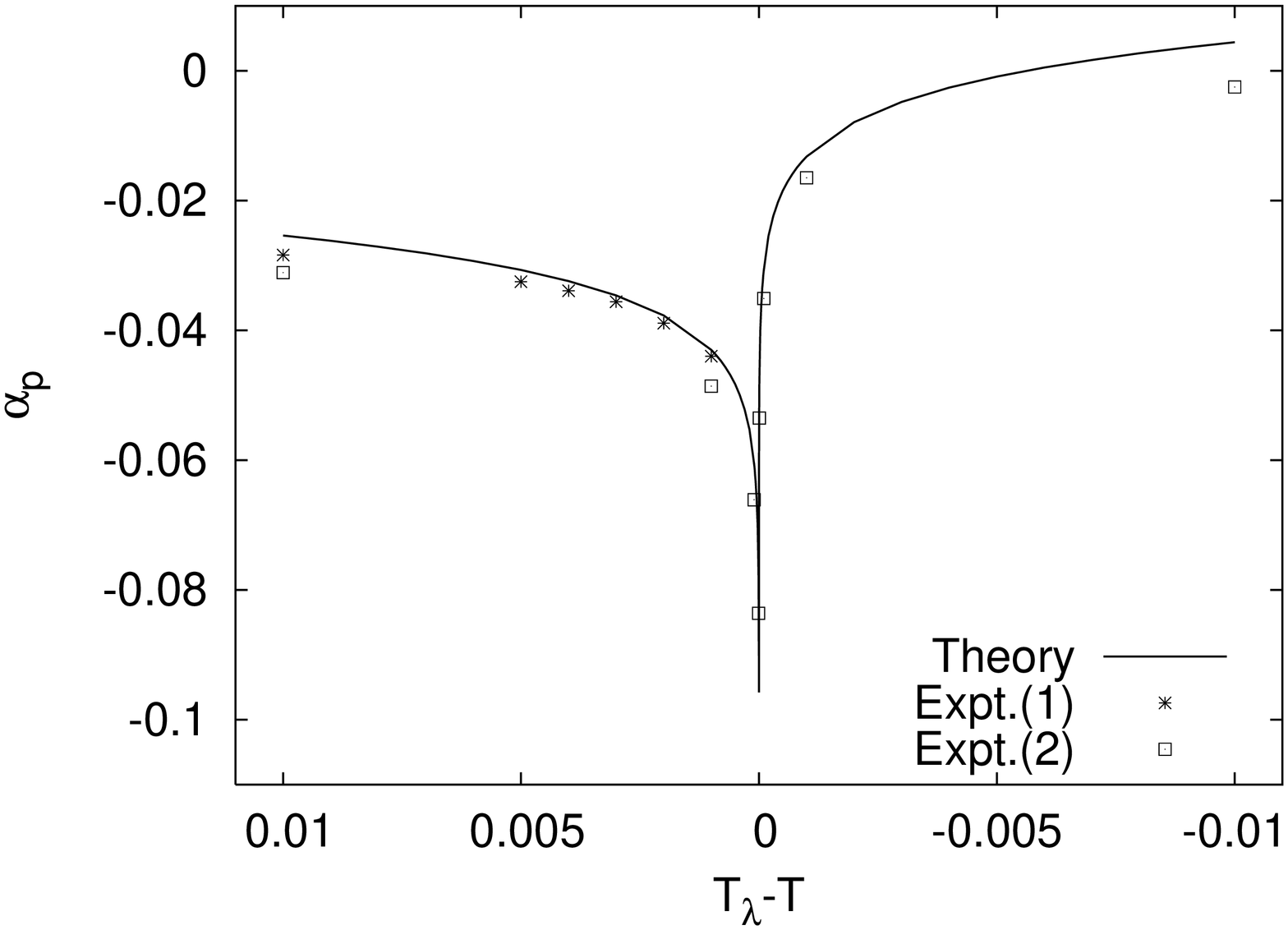}

\bigskip
\noindent
\centerline {Fig. 4 : $\alpha_p$ $vs.$ $T_{\lambda}-T$ curve for 
liquid $^4He$ around $\lambda-$point.}
\end{figure} 

\subsection{Isothermal Compressibility ($\kappa_T$):}
In what follows from [2], the isothermal compressibility is related 
to $\alpha_P$ and $C_p$, respectively, through
\begin{equation}
\alpha_P = V^{-1}\left(\frac{\partial V}{\partial T}\right)_t + 
\left( \frac{\partial P}{\partial T}\right)_t\kappa_T
\end{equation}
\begin{equation}
\kappa_T = \left[VT\left(\frac{\partial P}{\partial T}\right)_t^2
\right]^{-1}C_p- \frac{1}{V} \left[\frac{\left(\partial S/\partial T
\right)_t}{\left(\partial P/\partial T\right)_t^2} + \frac{
\left(\partial V/\partial T\right)_t}{\left(\partial P/\partial T\right)_t}
\right]
\end{equation}
This relation gives
\begin{equation}
\kappa_T = A_\kappa\log{\left|T-T_\lambda\right|} + B_\kappa
\end{equation}
Using $\left(\partial V/\partial T\right)_\lambda = 43.82 cm^3/mole/K$ and 
other parameters as used above we find $A_\kappa =-0.0001$ and 
$B_\kappa = 0.0141$ for $T<T_\lambda$ and $A_\kappa =-0.0001 $ and 
$B_\kappa = 0.0138$ for $T>T_\lambda$.  Our calculated $\kappa_T$ 
is depicted in Fig 5 with the experimental values taken from [9].   

\bigskip
\begin{figure}
\includegraphics [angle = -90, width = 0.9\textwidth]{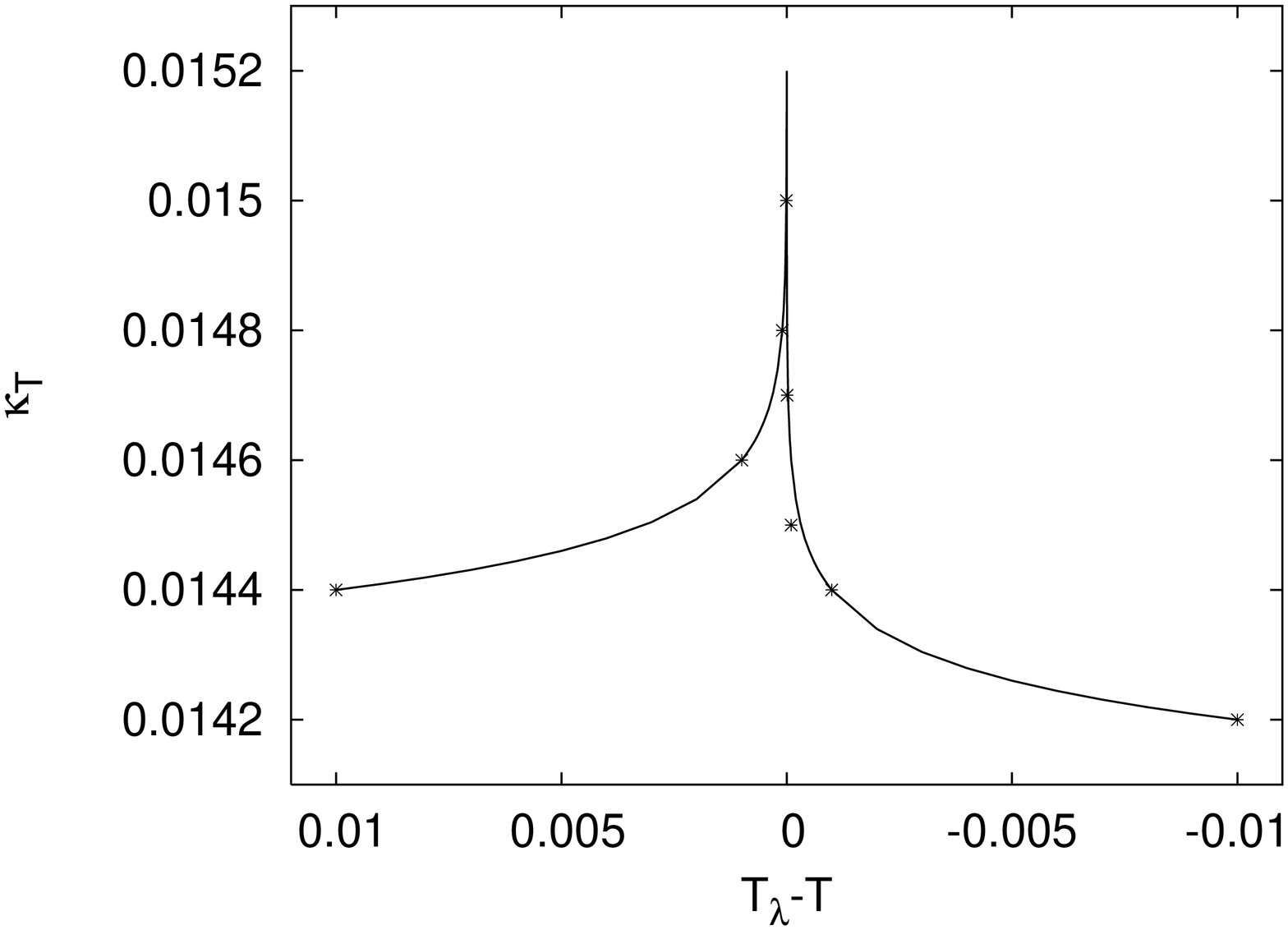}

\bigskip
\noindent
\centerline {Fig. 5 : $\kappa_T$ $vs.$ $T_{\lambda}-T$ curve for 
liquid $^4He$ around $\lambda-$point.}
\end{figure}  

\bigskip
\begin{figure}
\includegraphics [angle = -90, width = 0.9\textwidth]{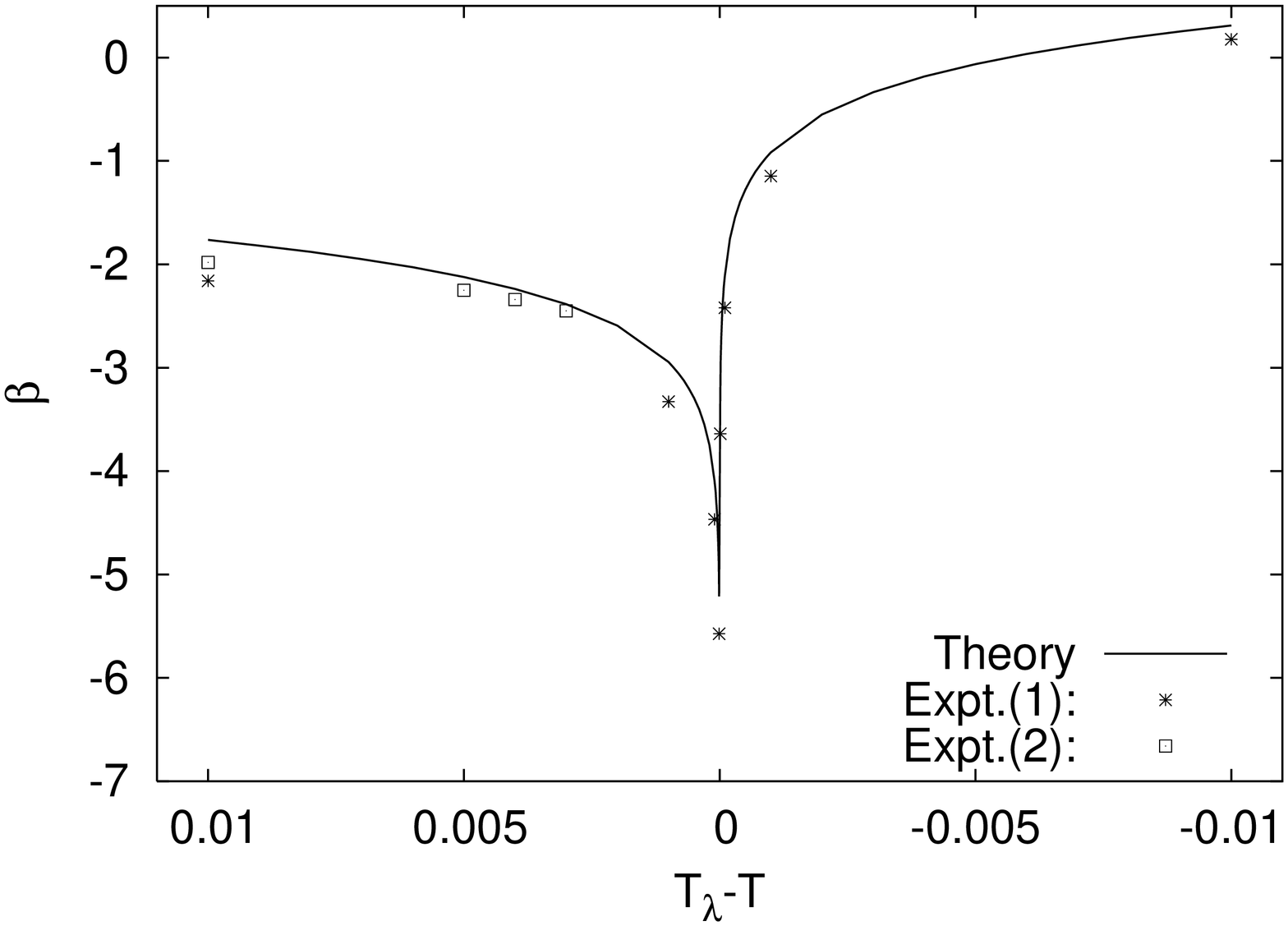}

\bigskip
\noindent
\centerline {Fig. 6 : $\beta$ $vs.$ $T_{\lambda}-T$ curve for liquid 
$^4He$ around $\lambda-$point.}
\end{figure}  
\subsection{Pressure Coefficient ($\beta$)}
From the thermodynamic relation
\begin{equation}
\left(\frac{\partial P}{\partial T}\right)_V = -\left(\frac{\partial P}
{\partial V}\right)_T\left(\frac{\partial V}{\partial T}\right)_P
\end{equation}
one can obtain
\begin {equation}
\beta = \left(\frac{\partial P}{\partial T}\right)_V = \frac{\alpha_P}
{\kappa_T}
\end{equation}
whose values at different $t$, obtained by using the calculated values 
of $\alpha_P$ and $\kappa_T$, are plotted in Fig 6.  The experimental 
points obtained, respectively, from [9] and [19] are marked as Expt.(1) 
and Expt.(2).  As expected, the divergence of $\beta$ at $\lambda-$point 
undoubtedly appears to be logarithmic.  
\section{Discussion}
Several experimental studies of specific heat of liquid $^4He$, {\it viz.} 
in bulk investigated out of earth's gravity [10,11] and confined to 
different geometries [12-16] have been performed near $\lambda-$point 
for various reasons discussed elegantly by these authors as well as by 
Bhattacharya and Bhattacherjee [38] and Schultka and Manousakis [39].   
In particular, these studies aim at verifying several predictions of 
renormalization group theory.  For the same 
reasons, extensive investigation of the $C_p$ of liquid $^4He$-$^3He$ 
mixtures near their $\lambda-$points have been also been made by 
Gasparini {\it et al} [40-41] and second sound propagation in liquid 
$^4He$ has been studied by Marek {\it et al} [42] and 
Swanson {\it et al} [43].  One, evidently, finds that the gravity of 
earth and the effects of finite size sample round off 
the $C_p$ of liquid $^4He$ at the $\lambda-$point and the position of 
its maximum is shifted to a temperature (say $T_m$) below $T_{\lambda}$ 
by a small amount.  However, in the absence of these effects $C_p$ 
divergence is logarithmic as concluded by Jain's theory [32].  In this 
context it may be mentioned that Jain's theory [32] does not incorporates 
weak effects arising from gravity of earth, finite size of the sample, 
{\it etc.} while concluding Eqn. 1.  Since such effects would always be 
present, the experimental $C_p$ would always be rounded for one reason 
or the other as one tries to reach closer and closer to $T_{\lambda}$.   

\bigskip
Jain's approach [32] to the microscopic theory of 
liquid $^4He$ type systems differs from conventional approaches [44] for its 
use of a macro-orbital (a kind of pair waveform [45]) to represent the 
self superposition state of a particle in a many body system.  One may find 
that each particle in such a system assumes self superposition state 
when its collision with other particle(s) forces its pre-collision and 
post-collision states (both represented, presumably, by plane waves of two 
different momenta) to have their superposition.  This, naturally, organizes 
two particles in collision at phase positions with $\Delta\phi = 2n\pi$, 
particularly, when particles are of low energy ({\it i.e.}, with wave 
length $\lambda$ equal to or larger than the inter-particle 
separation $d$) and this is achieved for nearly all particles in the system 
at low temperatures with thermal de Boroglie wave length $\lambda_T 
\approx 2d$ [32].  The fact, that Jain's approach gives due importance to 
all consequences of this superposition which represents {\it one of the 
simple truths} of wave nature of a quantum particle, ensures a better 
agreement of its predictions with experiments and this fact is successfully 
demonstrated by this paper.   In addition 
the merits of Jain's approach also rest with the facts that: (i) it makes no 
assumption like conventional theories of a system of interacting bosons 
which assume that $p=0$ condensate exists in superfluid state of liquid 
$^4He$ type systems or Cooper type bound pairs of $^3He$ atoms are the 
origin of superfluidity of liquid $^3He$ type systems, (ii) its all inferences 
are based on a systematic critical analysis of the solutions of 
$N-$particle Schr\"{o}dinger equation, (iii) as seen from [32, 46], it 
proposes a single framework for all many body systems of interacting 
bosons (or fermions) like $^4He$ (or electrons in solids or 
$^3He$) atoms, (iv) its simple 
mathematical framework is easy to comprehend and does not permit 
adjustable parameters like conventional theories where one finds widely 
different theoretical predictions depending on the values of such 
parameters (for example as discussed in [47], different papers using 
BCS picture predicted widely different superfluid 
$T_c$ for liquid $^3He$), (v) as shown in this paper it explains even those 
observations, {\it viz.} the logarithmic singularity of $C_p$ and related 
properties of liquid $^4He$, which found no basis in the framework of 
conventional theories, {\it etc.}   In addition, the fact that the 
theoretical results, obtained by using Jain's approach, for example for 
superfluid $T_c$ of liquid $^3He$ and its pressure dependence [47], 
excitation spectrum of liquid $^4He$ [48], and those reported here, match 
closely with experiments, clearly indicates that Jain's theoretical approach 
has great potential to explain the physics of widely different many body 
systems.  This has been demonstrated in his recent work related to the 
basic foundations of superconductivity [46] and ground state of $N$ hard 
core particles in 1-D box [49], unification of the physics of widely 
different many body systems [50] and wave mechanics of two hard core particles 
in 1-D system [45].  

\bigskip
Recently, Fliessbach [51] reported a semi-phenomenological microscopic model 
of $\lambda-$transition of $He-II$, proposed to be known as ``{\it almost 
ideal Bose gas model}'' (AIBG).  To this effect, he intuitively modifies 
the wave function $\Psi_{IBG} = \sum \Pi_k[\phi_k]^{n_k}$ representing a 
state of ideal Bose gas (IBG) with $\phi_k = \exp{(i{\bf k}.{\bf r}_j)}$ 
replaced by $\phi_k = \sin{(qx_j + \theta_j)}$ and introduces the concept 
of localized phase ordering by physical (or Dirichlet) boundary conditions 
at the walls of macroscopic volume $V$.   While there is considerable 
resemblance of these aspects of Fliessbach's single particle function  
$\phi_k = \sin{(qx_j + \theta_j)}$ with Jain's macro-orbital $\phi_{mo} = 
\sin{({\bf q}.{\bf r})}\exp{({\bf K}.{\bf R})}$ function which identifies 
each atom like a particle of quantum size $\lambda/2 = \pi/q$ moving freely 
with momentum ${\bf K}$, there is great deal of difference in the allowed 
values of $q$ permitted by his theory [51] ({\it viz.,} $q \ge \pi/L$ 
with $L$ being the size of macroscopic $V$) and Jain's theory [32] 
({\it viz.,} $q \ge \pi/d$ with $d = (V/N)^{1/3}$).  It may be noted 
Jain's theory uses purely microscopic considerations and the possibility 
of wave superposition to conclude a macro-orbital as the right wave 
function for a particle in a system like $He-II$ [32].  Interestingly, we note 
that Fliessbach finds that the energy, responsible for the 
logarithmic singularity of the specific heat of liquid $^4He$ at $\lambda-$ 
point, is related to phase ordering and it depends on $t$ as $t\ln{|t|}$ which 
agrees with Jain's result forming the basis of Eqn.1 used in the present 
analysis, of course with different multiplying factors to $t\ln{|t|}$.  
It may be noted that Fliessbach does not use his relation to show the 
nature of agreement of his theoretical results with experiments, hence 
no comparative analysis of his results with our results could be possible. 
However, since Fliessbach's model also uses phase ordering (although 
introduced intuitively) to explain the logarithmic nature of the said 
singularity, it may be argued that the phase ordering of particles, as 
concluded by Jain's microscopic theory, has strong foundation and the fact, 
that this ordering is an important characteristic of superfluid phase of 
liquid $^4He$ type systems, can not be ignored. In this context, it may 
be mentioned that studying the coupled low dimensional superfluids, 
Mathey {\it et. al.} [52] find that superfluids have strong tendency 
to phase lock which falls in line with an important inference of 
Jain's theory that inter-particle phase locking is an inherent 
aspect of a superfluid.

\section{Conclusion}
This study establishes that $C_p$ of liquid $^4He$, when observed in 
absence of the effects of earth's gravity or finite size of the sample, 
{\it etc.}, exhibits logarithmic singularity at $T_\lambda$ and this agrees 
closely with Jain's microscopic theory [32].  As shown in [32], this 
singularity is a consequence of order-disorder of particles in phase space 
forced by the wave nature of particles or quantum correlation between a pair 
of bosons.  We find that Jain's theory [32] is capable of accounting for 
the difference in $C_p$ at $T<T_\lambda$ and $T>T_\lambda$ for the same 
value of $\left|T_\lambda-T\right|$.  It is noted that $\alpha_P$, $\kappa_T$ 
and $\beta$  also exhibit logarithmic divergence for their relation with 
$C_p$ which implies that this effect also originates from the ordering of the 
particles in phase space.  The agreement of our theoretical values obtained 
from Jain's theory [32] with experimental results undoubtedly confirms the 
accuracy of Eqn. 1.  Evidently it may be concluded that Jain's 
macro-orbital theory is equipped to explain the origin of logarithmic 
divergence of thermodynamic response functions and it is more 
advantageous for its accuracy, simplicity and clarity.

\bigskip
\noindent
{\it Acknowledgement:} One of us (SC) is thankful to the Head, Department 
of Physics and Principal, St. Anthony's College for their 
encouragement to pursue research studies.

\bigskip
\noindent
{\bf References}

\bigskip
1. J. Wilks, The Properties of Liquid and Solid Helium, Clarendon Press, Oxford (1967).

2. G. Ahlers, in ``The Physics of Solid and Liquid Helium'', Part I [K.H. Bennemann and J.B. Katterson, eds.], pp 85-206, John Wiley and Sons, New York (1976).

3. W.H. Keessom and K.Clusius, Leiden Comm. 219e; Proc. Sect. Sci. K. ned. Akad. Wet. {\bf 35}, 307 (1932).

4. W.M. Fairbank, M.J. Buckinghum, and C.F. Kellers, in ``Proceedings of the Fifth International Conference on Low Temperature Physics and Chemistry'',[J.R. Dillinger, ed.], p. 50, University of Wisconsin Press, Madison, Wisc (1958).

5. C.F. Kellers, Ph.D. Thesis, Duke University, Durhum, N.C. (1960).

6. M.J. Buckinghum and W.M. Fairbank, in ``Progress in Low Temperature Physics'', [C.J. Gorter, ed.], Vol III, p. 80, North-Holland, Amsterdam (1961).

7. W.M. Fairbank and C.F. Kellers, in ``Critical Phenomena, Proceedings of a Conference'', [M.S. Green and J.V. Sengers, Natl. Bur.Std. Misc. Pub. No. 273, p. 71, U.S. GPO, Washington, D.C. (1966).

8. G. Ahlers, Phys. Rev. A, {\bf 3}, 696 (1971).

9. G.Ahlers, Phys.Rev. A, {\bf 8}, 530 (1973).

10. J. A. Lipa, D.R. Swanson, J.A. Nissen, T.C.P. Chui, U.E. Israelsson,
Phys. Rev. Lett., {\bf 76}, 944 (1996). 

11. J.A.Lipa, J.A. Nissen, D.A. Stricker, D.R. Swanson, and T.C.P. Chui, Phys. Rev. B, {\bf 68}, 174518-1 (2003).

12. J.A.Lipa and T.C.P. Chui, Phys. Rev. Lett., {\bf 51}, 2291 (1983).

13. M. Coleman and J.A.Lipa, Phys. Rev. Lett., {\bf 74}, 286 (1995).

14. J.A. Lipa, D.R. Swanson, J.A. Nissen, Z.K. Geng, P.R. Williamson, D.A. Stricker, T.C.P.Chui, U.E. Israelsson, and M.Larson, Phys. Rev. Lett., {\bf 84}, 4894 (2000).

15. T. Chen and F.M. Gasparini, Phys. Rev. Lett. {\bf 40}, 331 (1978).

16. S. Mehta and F.M. Gasparini, Phys. Rev. Lett., {\bf 78}, 2596 (1997).

17. H.K. Onnes and J.D.A. Boks, Leiden Comm. {\bf 170b} (1924).

18. E. Mathias, C.A. Crommelin, H.K. Onnes and J.C. Swallow, Leiden Comm.{\bf 172} (1925).

19. K.R. Atkins and M.H.Edwards, Phys. Rev., {\bf 97}, 1429 (1955).

20. E.C. Kerr., J. Chem. Phys., {\bf 26}, 292 (1957).

21. M.H. Edwards, Can. J. Phys., {\bf 36}, 884 (1058).

22. C.E. Chase, E. Maxwell and W.E. Millett, Physica,{\bf 27}, 1129 (1961).

23. E.C. Kerr and R.D. Taylor, Ann. Phys. {\bf 26}, 292 (1964).

24. O. Lounasmaa and L. Kaunisto, Ann. Acad. Sci. Fenn. 
Sev. A VI (Finland) No. 59 (1960).

25. O.V. Lounasmaa, Phys. Rev., {\bf 130}, 847 (1963).

26. H.A. Kiersted, Phys. Rev., {\bf 138}, A1594 (1965).

27. H.A. Kiersted, Phys. Rev., {\bf 153}, 258 (1967).

28. O.K. Rice, J.Chem. Phys., {\bf 22}, 1535 (1954).

29. A.B. Pippard, Phil. Mag.{\bf 1}, 473 (1956).

30. A.B. Pippard, The Elements of Classical Thermodynamics, Chap. IX, Cambridge Univ. Press (1957).

31. R.P. Feynman, {\it Satistical Mechanics}, W.A. Benjamin, Inc., Reading 
(1972), p.34. 

32. Y.S. Jain, J.Sci. Expl.,{\bf 16}, 77 (2002);  more detailed version 
of this paper is available for ready reference as a paper entitled, 
``{\it Macro-orbitals and microscopic theory of a system of interacting 
bosons}'' www.arxiv.org/cond-mat/0606571.  

33. M.E. Fisher, Reports on the Progress of Physics, {\bf 30}, 615 (1967).

34. B. Widom, J. Chem. Phys., {\bf 43}, 3892 (1965); {\bf 43}, 3898 (1965).

35. R.B. Griffiths, Phys. Rev.,{\bf 158}, 176 (1967).

36. L.P. Kadanoff, W.Gotze, D. Hamblen, R. Hecht, E.A. S. Lewis, V.V. 
Palciauskas, M. Rayl, and J. Swift, Rev. Mod. Phys., {\bf 39}, 395 (1967).

37. G. Ahlers, Phys. Rev. Lett., {\bf 23}, 464, 739 (1969).

38. S. Bhattacharya and J. K. Bhattacharjee, Phys. Rev. B {\bf 59}, 3341 
(1991).

39. N. Schultka and E. Manousakis, Phys. Rev. Lett. {\bf 75}, 2710 (1995). 

40. F.M. Gasparini and M.R. Moldover, Phys. Rev. B, {\bf 12}, 93 (1975).

41. F.M. Gasparini and A.A. Gaeta, Phys. Rev. B, {\bf 17}, 1466 (1978).

42. D. Marek, J.A. Lipa and D. Phillips, Phys. Rev. B, {\bf 38}, 4465(1988).

43. D.R. Swanson, T.C.P. Chui and J.A. Lipa, Phys. Rev. B, {\bf 46}, 
9041 (1992).

44 See Ref.[32] for a brief discussion on the two basic conventional 
approaches used to understand a system like liquid $^4He$. 

45. Y.S. Jain, Central Europ. J. Phys. {\bf 2}, 709 (2004).       

46. Y.S. Jain, {\it Basic foundations of microscopic theory of 
superconductivity}  www.arxiv.org/cond-mat/ 0603784

47. Y.S. Jain, {\it Superfluid $T_c$ of liquid helium-3 and its pressure 
dependence}, www.arxiv.org/cond-mat/0611298.

48. Y.S. Jain, {\it A study of elementary excitations of liquid helium-4 
using macro-orbital microscopic theory} www.arxiv.org/cond-mat/0609418.

49. Y.S. Jain, {\it Ground state of a system of N hard core quantum particles 
in 1-D box}  www.arxiv.org/cond-mat/0606409.

50. Y.S. Jain, J. Sc. Explor. {\bf 16}, 117 (2002).

51. T. Fliessbach, {\it A model for $\lambda-$transition of helium}, 
www.arxiv.org/cond-mat/0203353.

52.  L. Mathey, A. Polkovnikov and A.H. Castro Neto, {\it Phase locking 
transition of coupled low dimentional superfluids}, 
www.arxiv.org/cond-mat/0612174.

\end{document}